\documentclass[aip,reprint]{revtex4-1}
\usepackage{latexsym, amssymb, mathrsfs, graphicx}

\begin{document}
\title{Multi-chord fiber-coupled interferometry of supersonic plasma jets and comparisons with synthetic data} 

\thanks{Invited paper published as part of the Proceedings of the 19th Topical Conference on High-Temperature Plasma Diagnostics, Monterey, California, May, 2012.\\}

\author{Elizabeth C. Merritt}
\affiliation{Physics and Astronomy, University of New Mexico, Albuquerque, NM, 87131, USA}
\author{Alan G. Lynn}
\affiliation{Electrical and Computer Engineering, University of New Mexico, Albuquerque, NM, 87131, USA}
\author{Mark A. Gilmore}
\email{gilmore@ece.unm.edu}
\affiliation{Physics and Astronomy, University of New Mexico, Albuquerque, NM, 87131, USA}
\affiliation{Electrical and Computer Engineering, University of New Mexico, Albuquerque, NM, 87131, USA}
\author{Carsten Thoma}
\affiliation{Voss Scientific LLC, Albuquerque, NM, 87108, USA}
\author{John Loverich}
\affiliation{Tech-X Corporation, Boulder, CO, 80303, USA}
\author{Scott C. Hsu}
\affiliation{Physics Division, Los Alamos National Laboratory, Los Alamos, NM, 87545, USA}

\date{\today}

\begin{abstract}
A multi-chord fiber-coupled interferometer [Merritt et al., Rev. Sci. Instrum. $\bold{83}$, 033506 (2012)] is being used to make time-resolved density measurements of supersonic argon plasma jets on the Plasma Liner Experiment [Hsu et al., Bull. Amer. Phys. Soc. $\bold{56}$, 307 (2011)]. The long coherence length of the laser ($>$ 10 m) allows signal and reference path lengths to be mismatched by many meters without signal degradation, making for a greatly simplified optical layout. Measured interferometry phase shifts are consistent with a partially ionized plasma in which an initially positive phase shift becomes negative when the ionization fraction drops below a certain threshold. In this case, both free electrons and bound electrons in ions and neutral atoms contribute to the index of refraction. This paper illustrates how the interferometry data, aided by numerical modeling, are used to derive total jet density, jet propagation velocity ($\sim 15$--50 km/s), jet length ($\sim 20$--100 cm), and 3D expansion.
\end{abstract}

\maketitle

\section{Introduction}
Supersonic plasma jets are being studied on the Plasma Liner Experiment (PLX)\cite{PLX} as a method of forming imploding spherical plasma liners,\cite{Awe} which are a potential standoff driver for magneto-inertial fusion.\cite{Hsu} The plasma jets are formed and accelerated by railguns developed and built by HyperV Technologies and have initial densities on the order of $10^{17}$~cm$^{-3}$, radii of a few centimeters, lengths of a few tens of centimeters, and velocities up to 50~km/s.\cite{aps-witherspoon} An eight-chord, visible, heterodyne, fiber-coupled interferometer\cite{Merritt} has been designed and built for measuring the plasma density of a propagating single jet and the merging of multiple jets. In single jet propagation studies thus far, interferometry phase shift data show a positive value (due to free electrons) becoming negative (due to bound electrons associated with neutrals and ions) later in time, consistent with the ionization fraction $\mathit{f}$ dropping as the jet propagates and undergoes adiabatic expansion and radiative cooling. In such partially ionized plasmas, a two-color interferometer\cite{Huddlestone, weber} could simultaneously constrain $\mathit{f}$ and the ion plus neutral density $n_{tot}$. However, our single color system requires independent information on either $\mathit{f}$ or $n_{tot}$ to constrain the other.

The main purpose of the paper is to describe the interpretation of interferometer phase shift signals from partially ionized argon plasmas, to show the methodology of deducing plasma jet properties from the phase shift data, and to illustrate how synthetic interferometer data from plasma jet simulations can be used to help interpret and constrain the experimentally-deduced jet properties. More detailed physics results on the plasma jets relating to the PLX project will be reported elsewhere.  The remainder of the paper is organized as follows. Section~II very briefly describes the interferometer system.  Section~III derives the interferometer phase shift formula as a function of $\mathit{f}$ and $n_{tot}$ for argon plasmas.  Section~IV illustrates how several jet physical quantities are deduced from the phase shift data, and Sec.~V shows synthetic interferometer phase shift data that help constrain the interpretation of the experimental data.  Section~VI provides a summary.

\section{Interferometer design and setup}
The details of the eight-chord interferometer system are covered in a previous publication.\cite{Merritt} The interferometer uses a 320 mW, $\lambda_0 = 561$ nm laser with a 10 m coherence length, as specified by the manufacturer. The long coherence length of the laser allows for a sub-fringe phase resolution of $~4^o$ even with a path length mismatch between the probe and reference beams of a few tens of centimeters arising from the setup geometry. Figure 1 shows the experimental setup, in which the eight probe beams are positioned approximately perpendicularly to the direction of jet propagation (\emph{Z} axis), starting at \emph{Z} = 35 cm in 6.35 cm intervals.

\begin{figure}
  \includegraphics{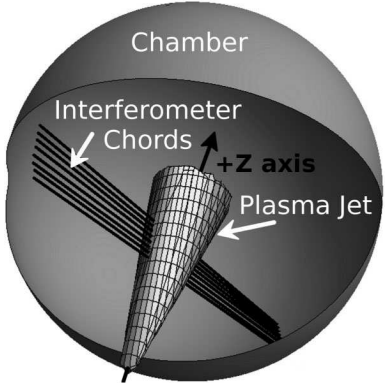}\\
  \caption{Schematic of the experiment in a 9 ft. diameter chamber. Eight interferometer chords each separated by 6.35 cm intersect the path of plasma jet propagation (\emph{Z} axis) as shown.  The plasma jet is launched at \emph{Z}=0, and the first chord is at \emph{Z}=35 cm.}
\end{figure}

\section{Phase Shift Analysis for a Partially Ionized Plasma}
For a strongly ionized plasma, the index of refraction for a laser beam of wavelength $\lambda_0$ can be approximated as a function of only the line-integrated electron density of the plasma:\cite{Hutchinson}
\begin{eqnarray}
\Delta \phi = \frac{e^2 \lambda_0}{4 \pi \epsilon_0 m_e c^2} \int n_e dl.
\end{eqnarray}
In this approximation it is assumed that the density of neutral particles in the plasma is negligible and that the frequency of the laser light has been chosen such that any interaction with the ions is also negligible. However, in a partially or weakly ionized plasma these assumptions are no longer valid and the ion and neutral particle densities may both contribute significantly to the index of refraction of the plasma.

\subsection{Neutral Atom and Ion Contributions to the Phase Change}
The index of refraction of a gas can be found using the Dale-Gladstone relation,\cite{alpher1}
\begin{eqnarray}
K\rho = (N-1) =\delta N,
\end{eqnarray}
where $\rho$ is the mass density of the gas, $K$ is the specific refractivity, $N$ is the refractive index, and $\delta N = N-1$. The specific refractivity of the gas is determined by the gas species and ionization state. For most neutral gases, the index of refraction and the density are well documented at standard temperature and pressure (STP) and can be used to calculate the specific refractivity. Thus the index of refraction of the neutral atoms in the plasma is:\cite{alpher1, kumar}
\begin{eqnarray}
\delta N_n = K m_n n_{n} =  \frac{\delta N_n^{STP}}{n_n^{STP}} n_{n}.
\end{eqnarray}
The index of refraction is related to the phase shift measured by the interferometer as follows:\cite{Hutchinson}
\begin{eqnarray}
\Delta \phi = \frac{\omega}{c} \int (N - 1) dl = \frac{2 \pi}{\lambda} \int \delta N dl,
\end{eqnarray}
where $\lambda$ is the wavelength of the laser. The phase change due to neutral atoms in the plasma is then: \cite{kumar}
\begin{eqnarray}
\Delta \phi_n = \frac{2 \pi}{\lambda}\frac{\delta N_n^{STP}}{n_n^{STP}} \int n_n dl.
\end{eqnarray}

Like the index of refraction of a neutral gas, the index of refraction for ions is also a result of the perturbation of the incident light by bound electrons in the ion. So, the Dale-Gladstone relation is still valid for the ions in the plasma. The phase change due to ions in the plasma is
\begin{eqnarray}
\Delta \phi_i = \frac{2 \pi}{\lambda}K_i m_i \int n_i dl.
\end{eqnarray}

\subsection{Expression for the Total Phase Shift}
The total interferometer phase shift is a superposition of the electron, ion, and neutral particle contributions since the indices of refraction are additive. \cite{Huddlestone} The interferometer signal analysis has been defined such that the contribution from any material with $N<1$ will give $\Delta \phi >0$, and materials with $N >1$ will give $\Delta \phi < 0 $. Since $N_e <1$ and $N_i, N_n > 1$, then the total phase shift is :
\begin{eqnarray}
\Delta \phi_{tot} &=& \Delta \phi_e - \Delta \phi_i - \Delta \phi_n \nonumber \\
&=&  \frac{\lambda e^2}{4 \pi \epsilon_0 m_e c^2} \int n_e dl - \frac{2 \pi}{\lambda} K_i m_i \int n_i dl \nonumber \\
&&- \frac{2 \pi}{\lambda}\frac{\delta N_n^{STP} }{n_n^{STP}} \int n_n dl.
\end{eqnarray}
Because the phase change from the ions and neutrals is in the opposite direction of the phase change from the electrons, using the approximation $\Delta \phi_{tot} = \Delta \phi_e$ can underestimate the actual electron density of the plasma.

The plasma jets are cold ($T_e \approx 1 eV$) by the time they intersect the interferometer chords, and thus the plasma is assumed to consist of only singly-ionized and neutral atoms, i.e., $n_e = n_i$. The total phase change is thus
\begin{eqnarray}
\Delta \phi_{tot} &=&( \frac{\lambda e^2}{4 \pi \epsilon_0 m_e c^2} - \frac{2 \pi}{\lambda} K_i m_i ) \int n_i dl \nonumber \\
&& - \frac{2 \pi}{\lambda}\frac{ \delta N_n^{STP} }{n_n^{STP}} \int n_n dl,
\end{eqnarray}
which is a function of only the line-integrated ion and neutral densities.

\subsection{Calculating the Total Atomic Density}
The total atomic density of the plasma is defined as the sum of the ion and neutral densities, $n_{tot} = n_i + n_n$. The interferometer is not equally sensitive to ion and neutral densities so a second relation between $n_i$ and $n_n$ is required in order to determine the total phase shift in terms of total atomic density. Assuming a value for the ionization fraction,
\begin{eqnarray}
\mathit{f} = \frac{n_i}{n_i + n_n} = \frac{n_i}{n_{tot}},
\end{eqnarray}
fulfills this requirement. The total phase shift can now be written in terms of $\mathit{f}$ and the total atomic density $n_{tot}$:
\begin{eqnarray}
\Delta \phi_{tot} &=&(\frac{\lambda e^2}{4 \pi \epsilon_0 m_e c^2} -  \frac{2 \pi}{ \lambda} K_i m_i + \frac{2 \pi }{\lambda}\frac{\delta N_n^{STP} }{n_n^{STP}}) \int \mathit{f} n_{tot} dl \nonumber \\
&&- \frac{2 \pi }{\lambda}\frac{\delta N_n^{STP} }{n_n^{STP}} \int n_{tot} dl.
\end{eqnarray}
Assuming an approximately uniform ionization along the path of the laser through the plasma, then the phase shift simplifies to
\begin{eqnarray}
\Delta \phi_{tot} &=& [( \frac{\lambda e^2}{4 \pi \epsilon_0 m_e c^2} - \frac{2 \pi}{ \lambda} K_i m_i) \mathit{f} \nonumber\\
&&- \frac{2 \pi }{\lambda}\frac{\delta N_n^{STP} }{n_n^{STP}}(1 -\mathit{f})] \int n_{tot} dl.
\end{eqnarray}

\subsection{Parameters with Argon}
\begin{figure}
  \includegraphics{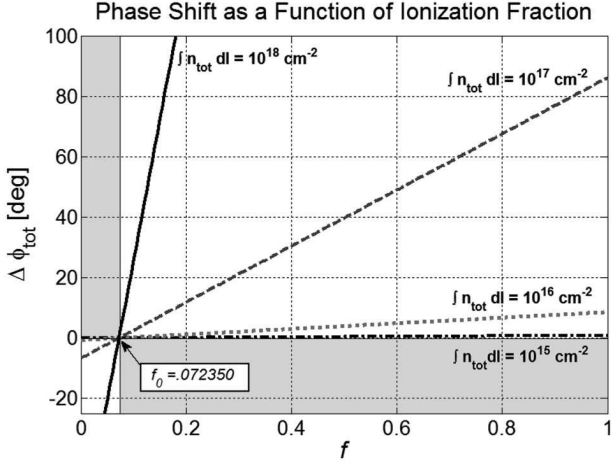}\\
  \caption{Calculate phase shift versus ionization fraction for an argon plasma at a variety of PLX-relevant atomic densities. The phase shift of $\Delta \phi_{tot} = 0$ occurs at $\mathit{f}_0 = 0.072350$. The gray areas are inaccessible parameters.}
\end{figure}

For argon, the Slater screening constant gives a specific refractivity for singly ionized argon:\cite{alpher2}
\begin{eqnarray}
K_{ArII} = 0.67 \times K_{ArI} = 0.67 \times \frac{\delta N_{ArI}}{m_{Ar} n_{ArI}}.
\end{eqnarray}
The phase shift from singly ionized argon is then
\begin{eqnarray}
\Delta \phi_{tot}&=&[ \frac{\lambda e^2}{4 \pi \epsilon_0 m_e c^2}\mathit{f} \nonumber\\
&&- (1 + 0.67 \mathit{f} - \mathit{f})\frac{2 \pi}{\lambda} \frac{\delta N_{ArI}^{STP} }{n_{ArI}^{STP}}] \int n_{tot} dl\\
&=& 1.6204 \times 10^{-17} [ \mathit{f} - 0.07235]\int n_{tot} dl,
\end{eqnarray}
a function of only $\mathit{f}$ and $n_{tot}$. For argon, $\Delta \phi_{tot} = 0$ occurs for $\mathit{f} = 0.07235$. The most general form for the phase shift is then
\begin{eqnarray}
\Delta \phi_{tot} = C [\mathit{f} - \mathit{f}_0] \int n_{tot}dl,
\end{eqnarray}
where $C$ is a scale factor for the gas species and $\mathit{f}_0$ is the ionization fraction for which $\Delta \phi_{tot} = 0$. Figure 1 shows a plot of $\Delta \phi_{tot}$ versus $\mathit{f}$ for a variety of $\int n_{tot} dl$ values of relevance for a single plasma jet. Since $\int n_{tot} dl$ is always positive, then $\Delta \phi > 0$ requires $\mathit{f} > \mathit{f}_0$ and $\Delta \phi< 0$ requires $\mathit{f} < \mathit{f}_0$.

\section{Experimentally Derivable Parameters}
\begin{figure*}
\includegraphics{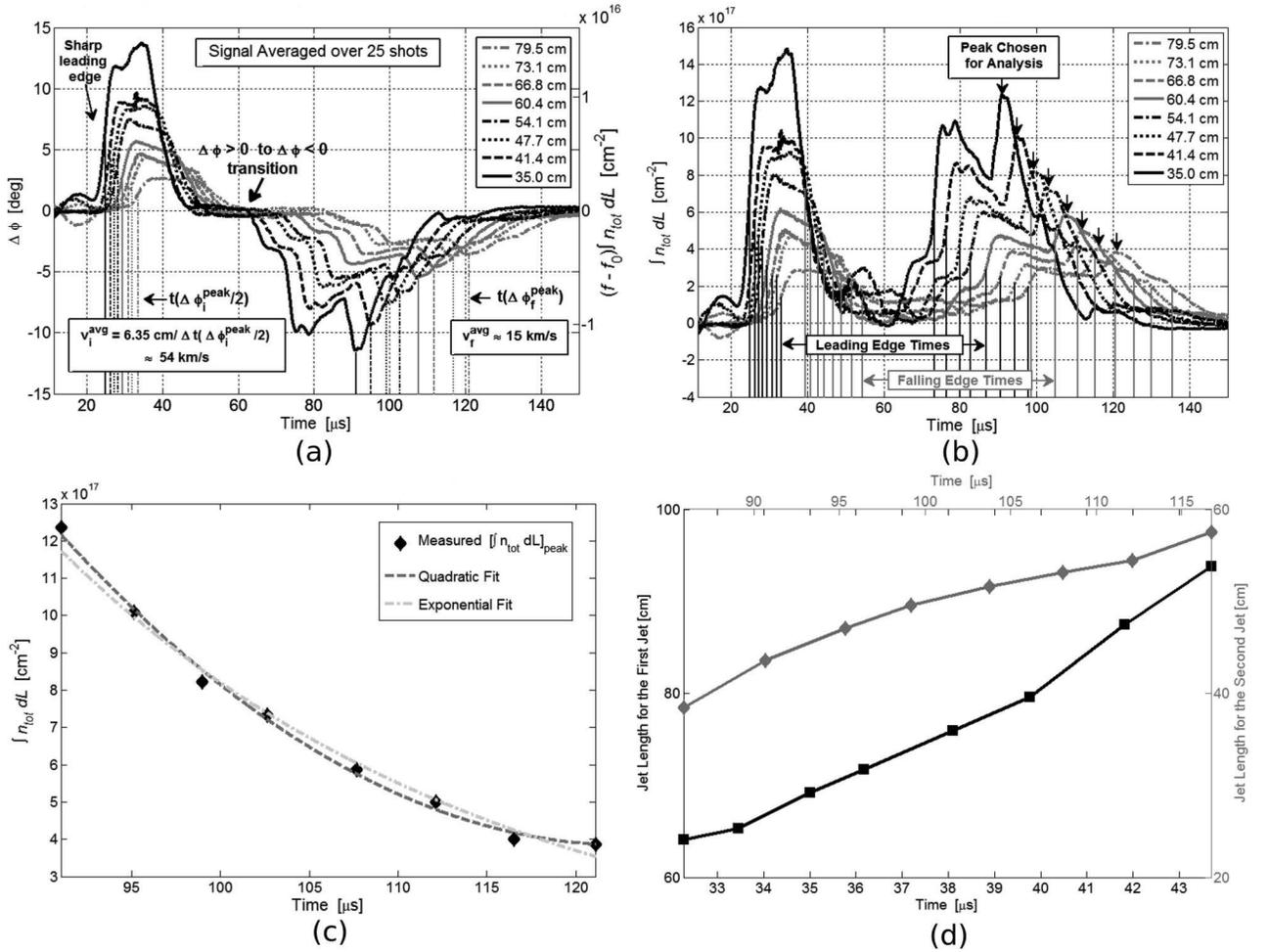}\\
\caption{(a) Phase shift versus time for the eight chords (distances from plasma gun indicated in legend). The $\mathit{f}$-dependent line-integrated density is given on the right-hand y-axis.  Arrival times of the leading edge and trailing peaks are marked as vertical lines. (b) Line-integrated ion plus neutral density versus time assuming $\mathit{f} = \mathit{f}_0 + 0.01$ for $\Delta \phi > 0$ and $\mathit{f} = \mathit{f}_0 - 0.01$ for $\Delta \phi <0$. Times of the leading and falling edges (half peak) are shown by vertical lines. (c) Peak line-integrated density versus time for the peaks indicated in (b). (d) Jet length versus time for the leading (left y-axis) and trailing jet (right y-axis) structures.}
\end{figure*}

PLX is currently studying argon single jet propagation from a HyperV Mark I plasma railgun \cite{aps-witherspoon} with $<350$ kA of peak current. Figure 3(a) shows the phase shift (left axis) versus time averaged over 25 shots with $I_{peak} = 260 \pm 5$ kA and a resulting chamber pressure rise after each shot of $P = 0.4 \pm 0.1$ mtorr. The signal for the averaged jet displays a sharp rise at the beginning of the signal, multiple peaks, and a transition from $\Delta \phi > 0$ to $\Delta \phi <0$.

\subsection{Ionization Information}
Since the phase shift is a function of both $\mathit{f}$ and $n_{tot}$, then independent information, such as from spectroscopy or synthetic diagnostics, is required to estimate $\mathit{f}$ and determine $\int n_{tot} dl$ (or vice versa). However, the sign of $\Delta \phi_{tot}$ alone can bound $\mathit{f}$. Any signal with $\Delta \phi_{tot} >0$ has a lower bound on the ionization fraction of $\mathit{f}_0$. Assuming a uniform $\mathit{f}$ along the path of the laser through the plasma, then according to Eq. (15) any signal with $\Delta \phi_{tot} <0$ has $\mathit{f} < \mathit{f}_0$. In Fig.3(a) the peak at 20 $\mu s < t <$ 60 $\mu s$ must have $\mathit{f} > 0.07235$ and all subsequent peak for  60 $\mu s < t <$ 150 $\mu s$ must have $\mathit{f} < 0.07235$. Thus the ionization is not constant along the jet's axial length. The ionization bounds set by the experimental signal can be used to calibrate simulated jet initial conditions such as initial jet temperatures, densities, and temperature and density distributions.

\subsection{Total Atomic Density and Jet Structure}
Once independent information about the ionization fraction ia available, then the interferometer can determine the total atomic density as well as several related jet parameters.

For purely illustrative purposes we assume an arbitrary ionization fraction of the form $\mathit{f}(t) = \mathit{f}_0 + a$ for $\Delta \phi > 0$ and $\mathit{f}(t) = \mathit{f}_0 - a$ for $\Delta \phi <0$, where $a =0.01$. This functional form was chosen because it is the simplest that fulfills the requirements that $ \mathit{f} > \mathit{f}_0$ for positive phase shifts and $\mathit{f} < \mathit{f}_0$ for negative phase shifts. The actual $\mathit{f}(t)$ for the jet is currently under investigation. Figure 3(b) is a plot of $\int n_{tot} dl$ versus time for the averaged jet in Fig. 3(a) assuming the arbitrary form of $\mathit{f}(t)$. The first distinguishable parameter is the order of magnitude of the line-integrated jet atomic density, $\int n_{tot} dl \approx 10^{18} cm^{-2}$ in Fig. 3(b). Higher jet densities are desirable for creating plasma liners that reach high pressure\cite{Awe} because the liner ram pressure scales linearly with liner density.

These jets have a consistent set of distinguishable structures: a sharp rise in atomic density that marks the leading edge of the jet, two distinct jet sections separated by a region of very low density, a single peak in the first jet section, and a double peak in the second jet section. The multiple peaks in the jet are possibly due to ringing of the gun electrical current.

\subsection{Jet Velocity}
Distinct structures in the signal have distinguishable arrival times at each interferometer chord. The distance between each chord is 6.35 cm, so the velocity of each distinguishable structure can be calculated using the time difference between the structure arrival times for consecutive interferometer chords. The jet velocity at the front of the jet is calculated using an arrival time of the leading edge calculated at half the peak value of the leading peak, $t = t(\Delta \phi^{peak}_i/2)$ as shown in Figs. 3(a)-(b). The leading edge of the jet has a velocity of $v^{avg}_i \approx 54$ km/s (averaged over 25 shots). The velocity for later peaks in the jet can also be calculated; the peak velocity for the final jet peak is calculated using the time difference between the peaks themselves for consecutive interferometer chords. The final peak velocity is $v^{avg}_f \approx 15$ km/s.

\subsection{Jet Expansion}
Phase shift data can be used to estimate jet radial and axial expansion as a function of time by invoking conservation of mass:
\begin{eqnarray}
n_{tot}(t)\times V(t) &=& constant \nonumber\\
(n_{tot} D)(t) \times D(t) \times L(t) &=& constant,
\end{eqnarray}
where V(t) is the volume of the jet,
\begin{eqnarray}
D(t) \propto \frac{1}{(n_{tot}D)(t)\times L(t)}
\end{eqnarray}
is the jet diameter, L(t) is the jet length, and $\int n_{tot} dL \approx (n_{tot}D)(t)$. Jet length is determined by
\begin{eqnarray}
L_{jet} = \frac{v^{avg}_f + v^{avg}_i}{2}(t_f - t_i).
\end{eqnarray}
where $i$ and $f$ denote the initial (leading) and final (falling) edges of each sub-jet, respectively, and the arrival times of the edges are taken at half the jet peak atomic density as indicated in Fig. 3(b). Jet length versus time for both sub-jets is plotted in Fig. 3(d). L(t) is non-constant for both sub-jets and implies axial expansion of the jet. The value of $(n_{tot}D)(t)$ for a sub-jet is determined from the maximum value and arrival time of a peak of the sub-jet. Fig. 3(c) plots $(\int n_{tot} dl)^{peak}(t)$ vs time for the peak indicated in Fig. 3(b) with quadratic and exponential fits to the data. These equations can be used to determine
\begin{eqnarray}
n_{tot}(t) = \frac{(n_{tot}D)(t)}{D(t)} \propto (n_{tot}D)^2(t) \times L(t).
\end{eqnarray}
in terms of experimentally-derived parameters.

\section{Synthetic Interferometry Data}
\begin{figure*}
\includegraphics{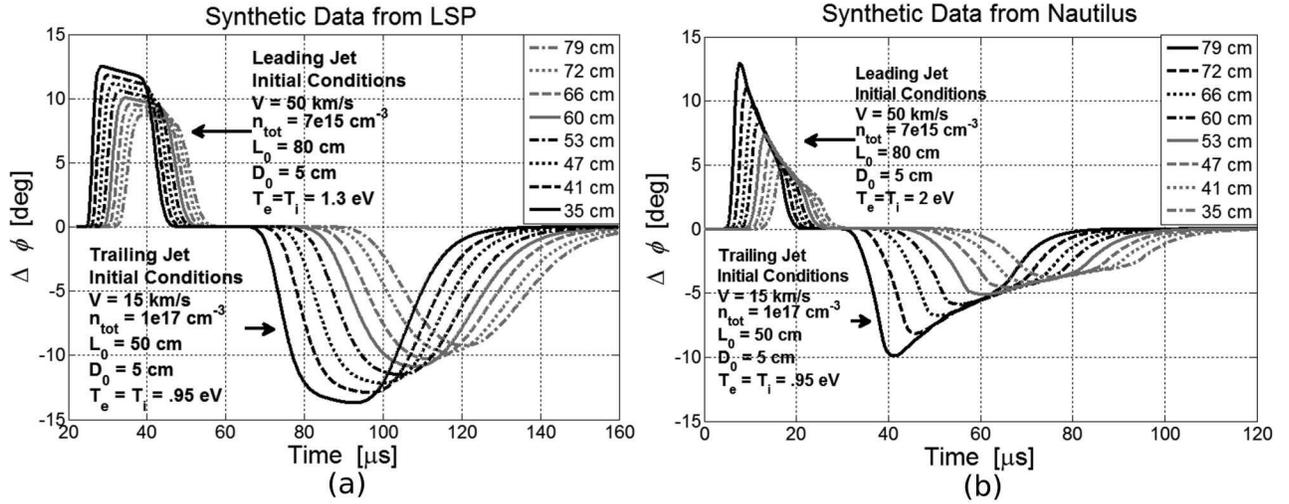}\\
\caption{Synthetic interferometer phase shift data from (left) LSP two-fluid and (right) Nautilus single-fluid jet simulations. Initial conditions used in the simulations are indicated in the plots. Synthetic chord positions relative to the plasma gun are given in the legends.}
\end{figure*}

For synthetic interferometry, the phase shift formula, Eq.(14), has been applied to simulated jet data from the single-fluid Nautilus \cite{loverich} and two-fluid Large Plasma Simulation (LSP)\cite{thoma} codes. Chord locations for the synthetic interferometry match those in the experiment. Relative positive and negative phase shift amplitudes, jet structures such as the sharp leading jet edge and the transition between positive and negative phase shifts, and jet arrival times at the interferometer chords are all used to calibrate simulation conditions.

Figures 4(a) and 4(b) show synthetic interferometer phase shift signals from 1D LSP\cite{thoma} two-fluid simulations and 3D Nautilus\cite{loverich} single-fluid simulations, respectively. Both simulations include non-local-thermodynamic-equilibrium (non-LTE) equation-of-state (EOS) tabular models,\cite{macfarlane} and assume a leading and a trailing jet with initial conditions as given in Fig. 4. The 1D LSP simulation assumes that the radius of both jets expands thermally for its path length calculation. The trailing jet requires a lower temperature and corresponding $\mathit{f}$ to generate a negative instead of positive phase shift. For these conditions both the negative and positive signals have roughly equal amplitudes, and are in approximate agreement with the experimental signal (Fig. 2). The non-LTE EOS model used bounds the jet temperature at $T_e < 1$ eV  for $\mathit{f} \ll 1$. The leading edges of the synthetic phase shift data also show a fast rise consistent with the experimental data. These simulations, which have produced synthetic data in good agreement with the experimental data, will be used to interpret and better understand the physics of single jet propagation.

\section{Summary}
An eight-chord, fiber-coupled interferometer with a long coherence length laser \cite{Merritt} is being used to study supersonic argon plasma jets. The long coherence length allows mismatches between probe and reference beams, allowing the use of only one reference beam for all eight probe beams, and simpler reconfigurability of the chords for doing different experiments. In single jet propagation experiments, the interferometer phase shift shows a positive phase shift becoming negative later in time. This is shown to be consistent with the $\mathit{f}$ of the plasma jet falling while the jet propagates. We derive a general phase shift formula accounting for free electrons, neutral atoms, and singly ionized argon ions. The formula depends only on $n_{tot}$ and $\mathit{f}$. Examples are shown for how the interferometer data, with some assumptions or independent input from other diagnostics or numerical modeling, can be used to deduce jet physical properties such as $n_{tot}$, $\mathit{f}$, velocity, and jet expansion. Finally, synthetic interferometer phase shift data calculated using Eq. (14) from single- and two-fluid simulations of argon plasma jet propagation are shown, giving an example of how the synthetic data can help with interpretation of the experimental phase shift data. Physics conclusions of the plasma jet studies as they relate to the PLX project will be reported elsewhere.

\begin{acknowledgments}
The authors acknowledge John Dunn, Joshua Davis, Thomas Awe, Igor Golovkin, Joseph MacFarlane, Sarah Messer, and John Thompson for their contributions, and Paul Bellan's group at Caltech for first alerting us to the interpretation of our observed negative phase shifts as possibly due to neutral atoms. This work was supported by the Office of Fusion Energy Sciences of the U.S. Dept. of Energy.
\end{acknowledgments}

\end{document}